\documentclass{aa}  

\usepackage{graphicx}
\usepackage{txfonts}
\usepackage{placeins}
\usepackage{soul}
\begin{document}

 \title{Solar atmospheric gravity wave phase differences: insights from MANCHA simulations and observations}

\titlerunning{Gravity waves}

   \author{Hirdesh Kumar
          \inst{1,2},
          E. Khomenko\inst{1,2},
          Tobias Felipe\inst{1,2}, 
          C. Quintero Noda\inst{1,2}
          }

   \institute{Instituto de Astrofísica de Canarias, 38205 C/ Vía Láctea, s/n, La Laguna, Tenerife, Spain\\
              \email{hirdeshchauhan71@gmail.com}
         \and
             Departamento de Astrofísica, Universidad de La Laguna, 38205 La Laguna, Tenerife, Spain\\
             }


\abstract
{Phase difference and coherence diagnostics are commonly used to detect and characterize gravity waves. However, those measurements are impacted by opacity and radiative transfer effects, which are generally overlooked.}
{We investigate the robustness of gravity wave phase propagation signatures derived from velocity and temperature and examine how these signatures appear in spectroscopic observables. We further compare diagnostics from MANCHA hydrodynamic simulations with observations of the quiet-Sun photosphere.}
{We perform $k$--$\omega$ phase difference and coherence analyses using three dimensional radiative hydrodynamic simulations computed with the MANCHA code. Velocity--velocity and temperature--temperature diagnostics are obtained at constant geometrical heights and optical depths. Synthetic Fe I 6173 {\AA} profiles are used to derive bisector velocities, continuum and line-core intensities. The same spectroscopic diagnostics are applied to high-resolution quiet-Sun observations obtained with the 1-m Swedish Solar Telescope (SST).}
{Velocity--velocity diagnostics show similar phase differences and coherence for diagnostics sampled at constant geometrical heights and optical depths, as well as for bisector-derived velocities from synthetic and observed Fe I 6173 {\AA} line profile. This indicates that the characteristic gravity wave signatures in velocity-based diagnostics are relatively insensitive to whether the same atmosphere is sampled at constant geometrical height or optical depth. In contrast, temperature--temperature diagnostics show systematic differences between the two sampling approaches, with stronger and more extended gravity-wave signatures at fixed geometrical height. The synthetic continuum and line-core intensities retain the gravity wave signature, but their phase differences closely resemble those of the temperature diagnostics at  constant optical depth. Similar qualitative behaviour is found in the SST observations, with quantitative differences in phase and coherence.}
{These results indicate that Doppler velocity provides a relatively robust diagnostic of gravity wave phase propagation when the same atmosphere is sampled at constant geometrical height or optical depth, whereas temperature and intensity diagnostics are more sensitive to opacity and radiative-transfer effects. Intensity observations can identify gravity wave signatures, but their quantitative phase differences should be interpreted in terms of their effective formation regions when inferring wave propagation properties and energetics. Consequently, using intensity- or temperature-based phase differences to infer wave propagation speeds and estimate energy fluxes may lead to an overestimation compared with velocity-based diagnostics.}


   \maketitle
%

\section{Introduction}

The solar atmosphere hosts a wide variety of magnetohydrodynamic (MHD) waves, which are believed to play a key role in the heating and driving dynamics of the outer solar atmosphere \citep{2006ApJ...640.1153C, 2007Sci...318.1574D, 2011ApJ...735...65F, 2015SSRv..190..103J, 2015LRSP...12....6K, 2019ApJ...871..155R}. Due to their importance, these waves have been studied extensively over the past few decades using both observations and numerical simulations \citep{2010ApJ...722..131F, 2016ApJ...830L..17Z, 2019ApJ...871..155R, 2023JASTP.24706071K}. Depending on the properties of the background medium, the dominant restoring force, and their observational signatures, MHD waves are commonly classified into several categories, including acoustic waves, gravity waves, magnetoacoustic waves, Alfv\'en waves, and magnetically guided modes such as sausage and kink waves \citep{Priest2014}.

Among these, atmospheric gravity waves have received considerable attention since the first estimate of their energy flux by \cite{2008ApJ...681L.125S}, with their theoretical description and propagation in stratified atmospheres having been described in the classical works of \cite{1981ApJ...249..349M,1982ApJ...263..386M, 1984frh..book.....M}. This pioneering study initiated extensive observational and theoretical efforts aimed at understanding the dynamics and energetic importance of gravity waves in the solar atmosphere \citep{2010MNRAS.402..386N, 2011MNRAS.417.1162N, 2011A&A...532A.111K, 2016ApJ...828...88H, 2017ApJ...835..148V, 2019ApJ...872..166V, 2020A&A...633A.140V, 2021RSPTA.37900170F, 2021RSPTA.37900177V, 2021RSPTA.37900178C, 2023AdSpR..72.1898K, 2023ApJ...952...58V, 2025ApJ...992..201V}. These waves are generated when turbulent convection penetrates locally into a stably stratified atmosphere, where buoyancy acts as the restoring force \citep{Priest2014}. They are characterized by negative phase shifts while transporting energy upward through the solar atmosphere. The energy flux estimated by \cite{2008ApJ...681L.125S} was found to be large enough to compensate for the chromospheric radiative losses, highlighting the possible contribution of gravity waves to chromospheric heating. Another important result of that study was the reported influence of magnetic fields on the propagation of gravity waves, particularly the suppression of wave power in magnetized regions. It was suggested that this suppression may be caused by the conversion of atmospheric gravity waves into Alfv\'en waves in the presence of magnetic fields, given the abundance of Alfv\'en waves observed in the corona \citep{2007Sci...318.1574D}. This interpretation was later supported by the numerical studies of \cite{2010MNRAS.402..386N, 2011MNRAS.417.1162N}, which showed that magnetic fields of the order of 10 G, especially when inclined, can strongly affect gravity wave propagation and promote mode conversion near the plasma $\beta = 1$ layer ($\beta$ is the ratio of gas pressure to magnetic pressure). Using CO5BOLD simulations, \cite{2017ApJ...835..148V} further investigated the influence of magnetic fields on gravity waves and identified signatures of wave scattering and suppression through comparisons of phase differences in quiet and magnetized regions.

Observational signatures consistent with the predicted influence of magnetic fields on atmospheric gravity waves through $k-\omega$ phase and coherence diagnostic diagrams were subsequently reported by \cite{2023AdSpR..72.1898K}. In this work, photospheric line-of-sight velocities were derived at two heights within the Fe I 6173 {\AA} spectral line using observations from the Interferometric BIdimensional Spectrometer (IBIS: \cite{2006SoPh..236..415C}) at the Dunn Solar Telescope (DST). The analysis was complemented by photospheric continuum intensity ($I_{\rm c}$) observations acquired from the Helioseismic and Magnetic Imager \citep[HMI;][]{2012SoPh..275..229S} and 1700 {\AA} and 1600 {\AA} intensity observations obtained from the Atmospheric Imaging Assembly (AIA: \cite{2012SoPh..275...17L}), both onboard the Solar Dynamics Observatory (SDO; \cite{2012SoPh..275....3P}) spacecraft. The analysis revealed reduced coherence in magnetized regions together with a corresponding suppression of atmospheric gravity waves. In addition, a comparison of $k-\omega$ phase-difference diagrams obtained at disk center and slightly off disk center showed enhanced negative phase shifts over an extended wavenumber range, suggesting possible detection of obliquely propagating gravity waves. More recently, \cite{2025ApJ...992..201V} investigated gravity waves under different magnetic-field configurations and also reported signatures of gravity-wave scattering and suppression in weighted coherence and phase-shift analyses availing the observations acquired from the IBIS and SDO solar facilities. However, negative phase signatures were absent from the corresponding $k$--$\omega$ phase-difference diagrams. Additionally, comparisons between observations at disk center and away from disk center revealed no clear evidence of oblique gravity wave propagation, possibly indicating that the excitation of these waves is inherently stochastic.

Despite the significant progress made in understanding solar atmospheric gravity waves, particularly their characteristics under different projection angles and magnetic-field configurations, an important question remains unresolved: how reliably do commonly used observables represent the intrinsic properties of gravity waves? Doppler velocities and intensity measurements are widely employed to study wave propagation, yet they sample the atmosphere differently and are expected to respond differently to opacity and radiative-transfer effects. Consequently, it is still unclear to what extent the phase differences inferred from these observables reflect the true wave propagation or are modified by the formation properties of the diagnostic itself.

In this work, we investigate the reliability of velocity and intensity diagnostics of atmospheric gravity waves by combining realistic hydrodynamic (HD) simulations computed with the MANCHA code \citep{2006ApJ...653..739K,2010ApJ...719..357F,2024SoPh..299...23M}, synthetic Fe I 6173 {\AA} spectral profiles, and observations obtained from the 1-m Swedish Solar Telescope (SST: \cite{Scharmer2003SST}). We first compare wave properties derived at constant geometrical heights and constant optical depths to isolate the influence of opacity on gravity waves. We then examine how these effects propagate into synthetic observables and assess whether they are consistent with observations. This approach enables us to identify which observables provide the most reliable measurements of gravity-wave phase propagation and to place important constraints on the interpretation of intensity-based diagnostics and, thus, the estimation of the energetics of gravity waves. The paper is organized as follows. Section 2 describes the numerical simulations and observations, and Section 3 presents the analysis and results. In Section 4, we discuss the implications of our findings and summarize the main conclusions.


\section{Numerical Simulations and Observations}
\label{sec2}

\subsection{MANCHA Simulations}

The numerical simulations analyzed in this work were performed using the three-dimensional radiative magnetohydrodynamic code MANCHA3D, which has been extensively developed to study wave propagation, magnetoconvection, and other dynamical processes in the solar atmosphere. The code solves the time-dependent MHD equations on a Cartesian grid and incorporates a realistic equation of state and radiative transfer, allowing realistic simulations of the solar photosphere and chromosphere. Detailed descriptions of the numerical algorithms and the physical modules implemented in the code are provided by \cite{2006ApJ...653..739K, 2010ApJ...719..357F, 2024SoPh..299...23M}.

The simulation employed in this study corresponds to a realistic radiative HD model of the quiet-Sun atmosphere, in which convective motions self-consistently excite acoustic and gravity waves that propagate through the photosphere and chromosphere. The three-dimensional computational domain contains 768 $\times$ 768 $\times$ 138 grid points in the x, y and z-directions, respectively. A uniform grid spacing of 32 km was adopted in all three
directions, corresponding to a physical domain size of approximately 24.6 $\times$ 24.6 $\times$ 4.4 Mm$^3$ . The bottom boundary is placed at z=-2.8 Mm, with z = 0 defined as the average height where $\log\tau=0$. The upper boundary is located at z=1.6 Mm.

\begin{figure}
	\centering
	\includegraphics[width=0.99\linewidth]{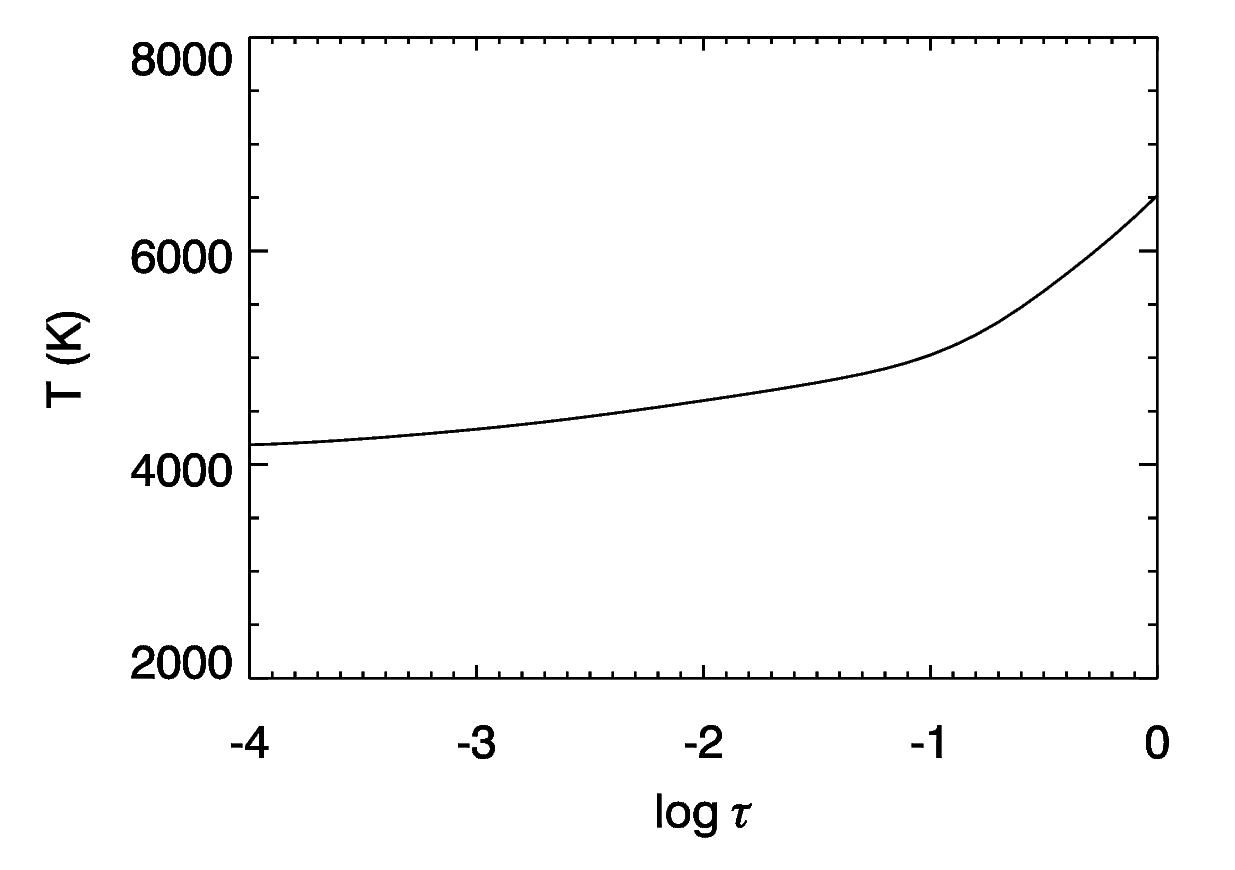}
	\caption{Temperature variation obtained by averaging over the full field of view  of the MANCHA snapshot.}
	\label{mancha_temp_variation}
\end{figure}

The simulations were evolved for a total time duration of 4 h, and the outputs were saved every 10 s. For the present analysis, velocity and temperature snapshots sampled at a cadence of 50 s were utilized. Figure \ref{mancha_temp_variation} shows the temperature variation as a function of height, estimated by taking the mean temperature over the full computational domain at a randomly selected time step. The spatial and temporal resolutions adopted are sufficient to resolve the gravity wave spectrum.  The 50 s temporal sampling cadence corresponds to a Nyquist frequency $(\nu_{\rm NY})$ of 10 mHz. The frequency resolution $(\Delta \nu)$, determined by the total duration of the analyzed time series, is 69.4 $\mu$Hz. The horizontal domain size provides a horizontal wavenumber resolution $(\Delta k_h)$ of approximately 0.255 rad Mm$^{-1}$, while the spatial sampling corresponds to a Nyquist wavenumber $(k_{\rm NY})$ of 98.1 rad Mm$^{-1}$.

\subsection{Synthetic Fe I 6173 {\AA} spectra}

The synthetic Fe I 6173 {\AA} Stokes profiles were computed using the Stokes Inversion based on Response functions (SIR) code \citep{1992ApJ...398..375R} in its forward synthesis mode under the assumption of local thermodynamic equilibrium (LTE). Since atmospheric stratification from the MANCHA simulations is defined on a geometric height scale, each simulation snapshot was first converted to a logarithmic continuum optical depth scale ($\log\tau$) using the \texttt{modelador} utility distributed with the SIR package. This conversion was carried out by computing the continuum opacity at each atmospheric layer from the local thermodynamic variables and integrating the radiative transfer relation,
\[
d\tau_\lambda = -\kappa_\lambda \rho\,dz,
\]
where $\kappa_\lambda$ is the continuum mass absorption coefficient, $\rho$ is the mass density, and $z$ is the geometrical height. The resulting atmospheric quantities were subsequently interpolated onto a regular grid $\log\tau$, which served as an input atmosphere for the radiative transfer calculations. The Fe I 6173 {\AA} Stokes profiles were synthesized with a spectral sampling of 10 m{\AA}, without including the micro and macro turbulence. To facilitate a direct comparison with observations obtained using the CRisp Imaging SpectroPolarimeter \citep[CRISP;][]{2008ApJ...689L..69S} at the SST, the synthesized spectra were spectrally degraded by convolving them with a Gaussian point spread function having a full width at half maximum (FWHM) of 50 m{\AA}, closely matching the spectral transmission profile of the CRISP instrument. The degraded spectra were subsequently resampled to a uniform spectral sampling of 30 m{\AA}. Although our SST observations have a spectral sampling of 35 m{\AA}, the small difference in wavelength sampling would have a negligible impact on the analysis, allowing a consistent comparison between the synthetic and observed spectra.

\subsection{Bisectors and Fe I 6173 {\AA} line sensitivity}

\begin{figure*}
    \centering
    \includegraphics[width=0.45\linewidth]{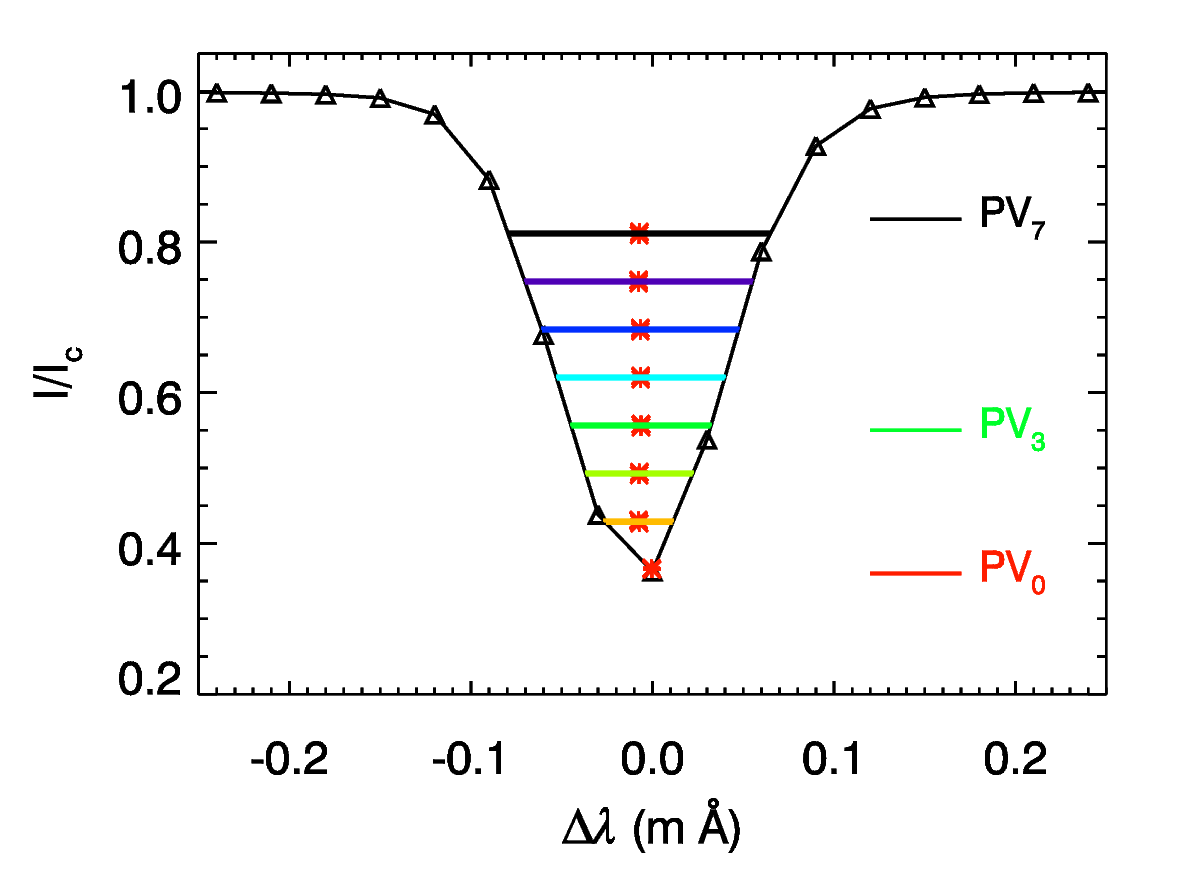}
    \includegraphics[width=0.45\linewidth]{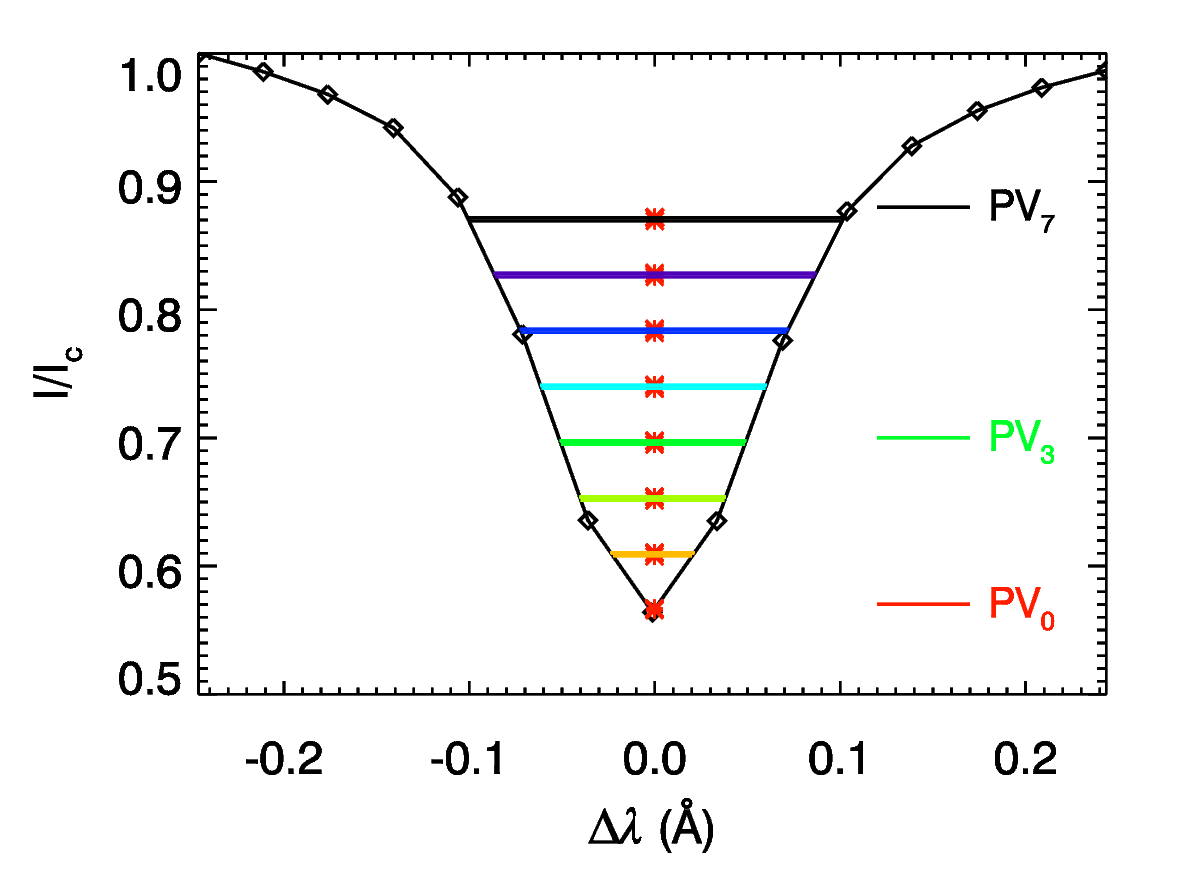}
\caption{Photospheric Fe I 6173 {\AA} spectral profiles obtained by averaging over the full field of view of the MANCHA simulation and the SST observations. The left panel shows the synthetic Fe I 6173 {\AA} profile from the MANCHA simulation, sampled at 30 m{\AA}, while the right panel shows the observed profile from quiet-Sun region obtained with the SST, sampled at 35 m{\AA}. The bisector points are defined at intervals of 10$\%$ of the line-depth intensity.}
\label{Fe_6173_bisector_profile_SST_Mancha}
\end{figure*}

\begin{figure*}
	\centering
	\includegraphics[width=0.9\linewidth]{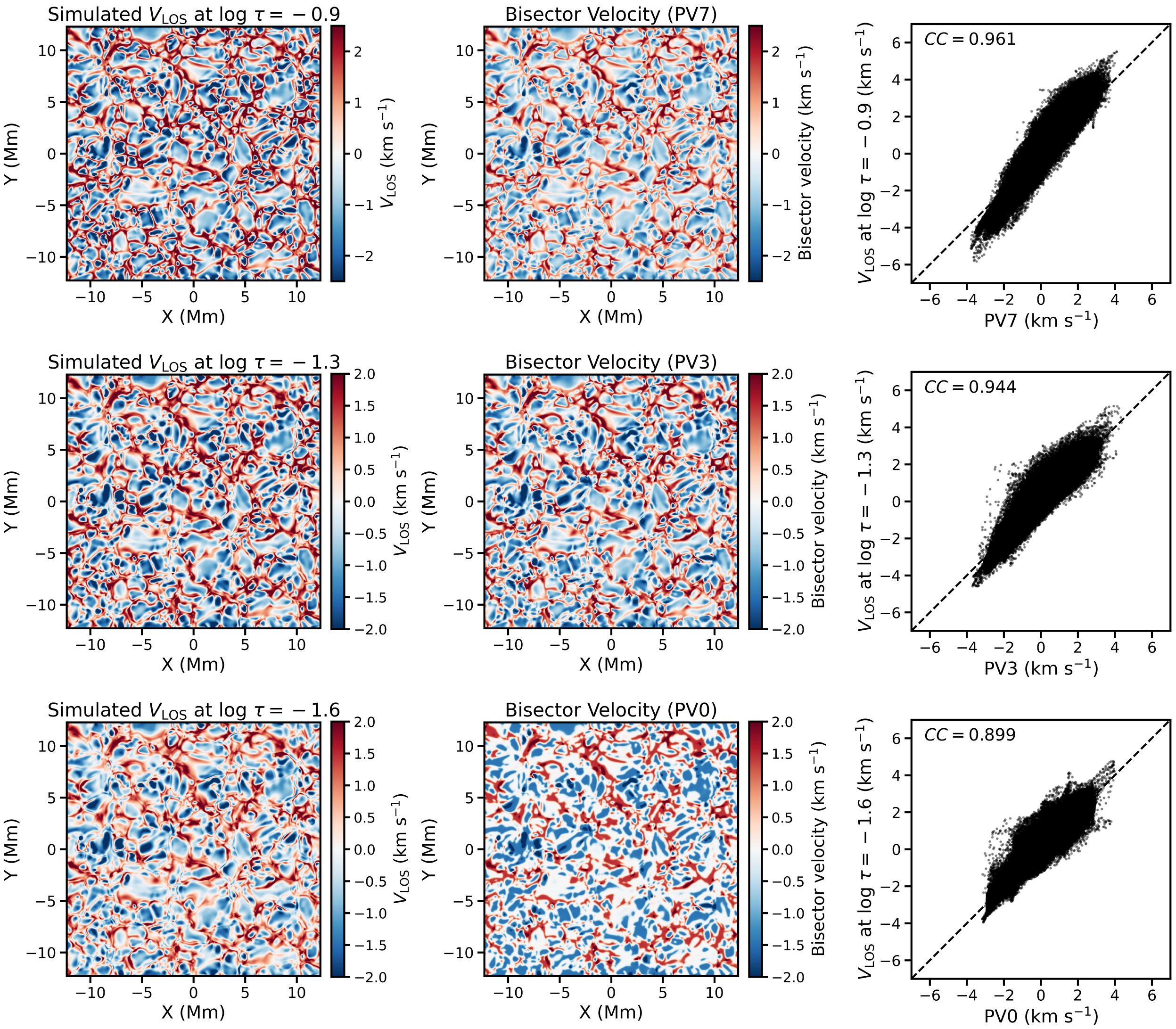}
\caption{The left panels show the line-of-sight (LOS) velocities from the MANCHA simulation at $\log \tau=-0.9$, $-1.3$, and $-1.6$. The middle panels show the corresponding LOS velocities derived by applying the bisector method to the synthetic Fe I 6173 {\AA} spectral line at the PV7, PV3, and PV0 levels, respectively (see the left panel of Figure \ref{Fe_6173_bisector_profile_SST_Mancha}). The right panels show the highest cross-correlation between the simulated and bisector-derived velocities at the corresponding optical-depth levels, illustrating the correlation between the two velocity measurements.}
\label{Bis_optical_depth_cc}
\end{figure*}

Multi-height line-of-sight (LOS) velocities are estimated using the bisector method applied to the photospheric Fe I 6173 {\AA} spectral line. Figure \ref{Fe_6173_bisector_profile_SST_Mancha} shows the synthetic and observed Fe I 6173 {\AA} spectral profiles, obtained by averaging the intensity over the full field of view (FOV) from the MANCHA snapshot and SST observations, respectively. The bisector method is applied at intervals of 10 $\%$ line-depth intensity. We refer to them as PV0--PV7, where PV0 corresponds to the Doppler shift of the line core and the subsequent labels from PV1 to PV7 are given to bisectors between 10$\%$ and 70$\%$ line-depth intensity. We further assess the sensitivity of the bisector-derived multi-height LOS velocities extracted from the synthetic Fe I 6173 {\AA} profile by comparing them with the MANCHA simulation. Figure \ref{Bis_optical_depth_cc} shows the simulated velocities versus the bisector-derived velocities, along with their peak correlation. We find that the PV7, PV3, and PV0 velocity levels sample approximately $\log \tau = -0.9$, $-1.3$, and $-1.6$, respectively, corresponding to mean geometrical heights of approximately 39 km, 101 km, and 144 km above $z = 0$.

\subsection{SST Observations}

Quiet-Sun observations acquired near the solar disc center on 24 April 2019 with the SST telescope are also used in this study to compare the inferences obtained from the simulations. Photosphere was observed in the Fe I 6173 {\AA} spectral line, sampled at 15 wavelength positions with a spectral spacing of 35 m{\AA}. This  observing sequence covers the interval from 08:30 to 10:30 UT, with a temporal cadence of 28.18 s and a spatial sampling of 0.0591 arcsec pixel$^{-1}$. The observations benefited from the SST adaptive optics system \citep{2019A&A...621A..68S} and were further restored using the Multi-Object Multi-Frame Blind Deconvolution (MOMFBD) technique \citep{2005SoPh..228..191V}. Figure \ref{Sample_maps_Ic_SST_appendix} shows the photospheric continuum intensity map, which illustrate selected field-of-view for the current analysis. This data set is publically available at the SST site: \url{https://dubshen.astro.su.se/sst_archive/search}.

\section{Analysis and Results}
\label{Sect3}

\begin{figure*}
	\centering
	\includegraphics[width=0.99\linewidth]{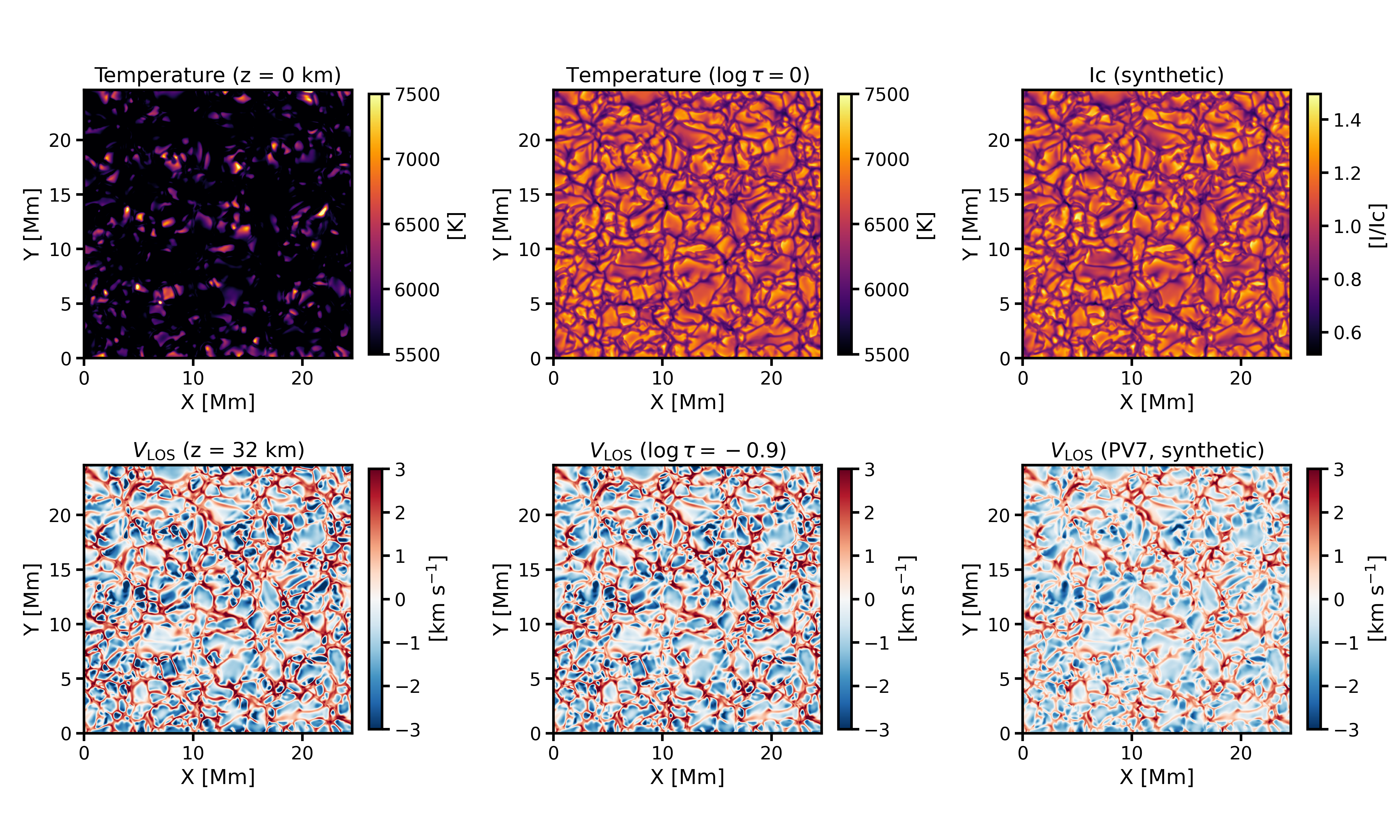}
\caption{Representative maps of temperature (top row) and line-of-sight (LOS) velocity (bottom row) from the MANCHA simulation, along with the synthetic continuum intensity and bisector-derived LOS velocity at the PV7 level from the synthetic Fe I 6173 {\AA} line. The geometrical-height and optical-depth maps are obtained by sampling the same simulation snapshot at the corresponding atmospheric levels.}
	\label{Sample_maps_MANCHA_Sim}
\end{figure*}

\begin{figure}
	\centering
	\includegraphics[width=0.99\linewidth]{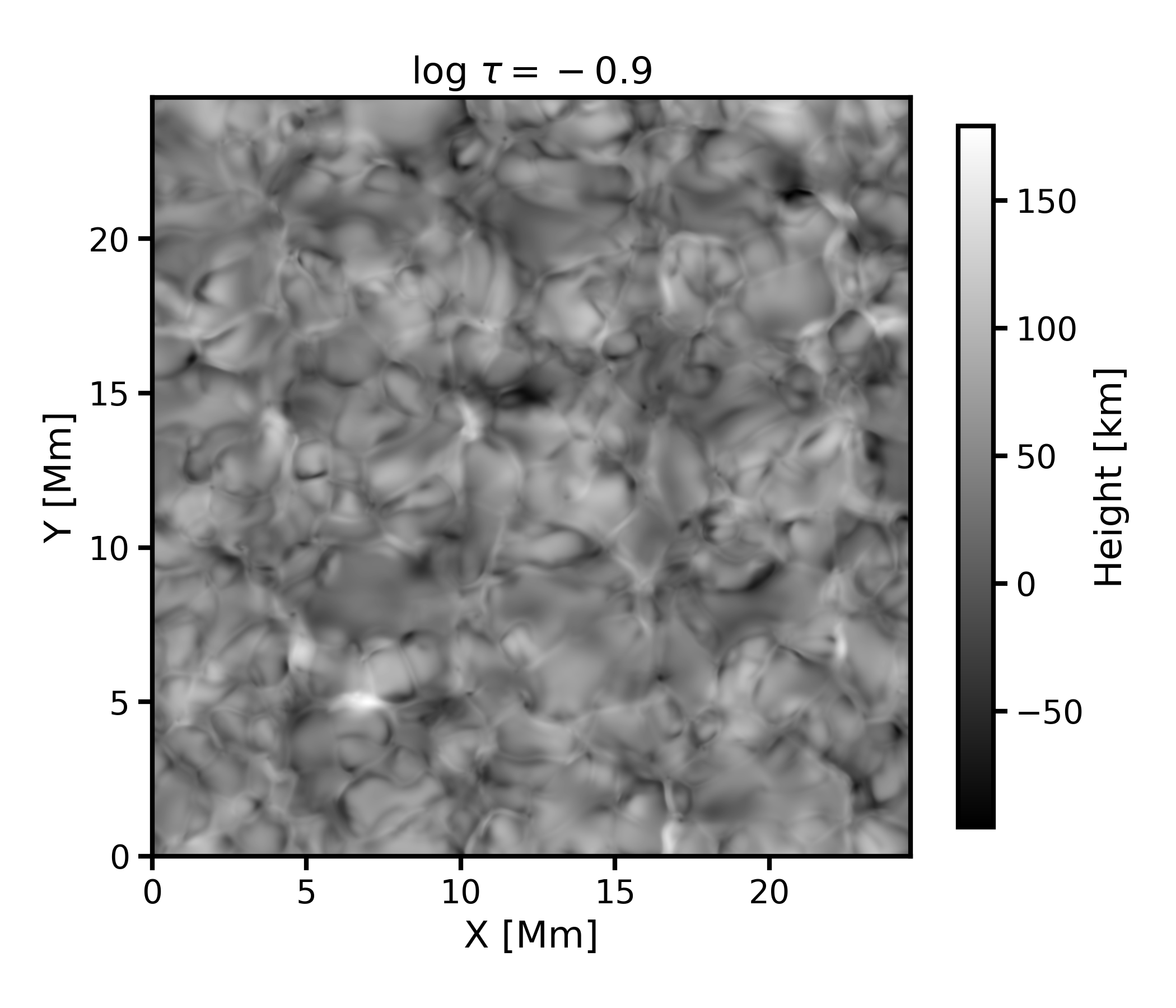}
	\caption{Map of the geometrical height corresponding to the optical-depth level $\log\tau=-0.9$ in the MANCHA simulation at a randomly selected time step.}
	\label{Sample_maps_logtau}
\end{figure}

The top panels of Figure \ref{Sample_maps_MANCHA_Sim} shows the temperature maps obtained from the MANCHA simulation at the geometrical height $z = 0$ km and optical depth $\log \tau = 0$, respectively, along with the corresponding synthetic continuum intensity map. The bottom panels show the line-of-sight (LOS) velocity maps obtained from the simulation at the geometrical height $z = 32$ km and optical depth $\log \tau = -0.9$, respectively. For comparison, the bisector-derived velocity at the PV7 level, obtained from the synthetic Fe I 6173 {\AA} line, is also shown. The PV7 level corresponds to $\log \tau = -0.9$, which is associated with a mean geometrical height of approximately $39$ km, allowing for a direct comparison with the velocity map sampled at the corresponding optical-depth level. Figure \ref{Sample_maps_logtau} shows the variation of the geometrical height corresponding to a constant optical-depth level of $\log\tau=-0.9$. This optical-depth layer corresponds to a mean geometrical height of approximately 39 km above $z=0$. However, at this randomly selected time step, the geometrical height of this optical-depth surface varies substantially across the simulation domain, spanning a range of more than 250 km.\\

We analyze cross-spectral properties, namely, the phase difference ($\delta \phi (k_h, \nu)$) and coherence ($C (k_h, \nu)$) diagnostic diagrams, derived from observables that sample the lower solar atmosphere, to examine the propagation characteristics of gravity waves. We have used pairs of temperature--temperature, intensity--intensity, and velocity--velocity diagnostics sampled at two atmospheric heights. Their cross-spectra are derived as functions of horizontal wavenumber ($k_{h}$) and frequency ($\nu$) \citep[][]{2017ApJ...835..148V,2023AdSpR..72.1898K}. These diagnostics are constructed by sampling the same atmospheric model at pairs of constant geometrical heights or constant optical depths, as appropriate, and by using bisector-derived velocities from the synthetic and observed Fe I 6173 {\AA} profiles.

\noindent For two observables, $f_1(x,y,t)$ and $f_2(x,y,t)$, the complex cross-spectrum is computed from their three-dimensional Fourier transforms as follows:

\begin{equation}
X_{12}(k,\nu)=f_1(k,\nu) \times f_2^{*}(k,\nu),
\end{equation}

\noindent where the asterisk denotes complex conjugation, $k=(k_x,k_y)$ is the horizontal wave number, $\nu$ is the cyclic frequency, and $\times$ indicates multiplication. The indices 1 and 2 refer to measurements obtained at heights $z_1$ and $z_2$, respectively.

\noindent The phase difference between the two atmospheric layers is obtained from the argument of the complex cross-spectrum,

\begin{equation}
\delta\phi(k,\nu)=\arg\left[X_{12}(k,\nu)\right].
\end{equation}

\noindent In addition, the coherence is evaluated from the normalized cross-spectral amplitude,

\begin{equation}
C(k,\nu)=\frac{\left|\left\langle X_{12}(k,\nu)\right\rangle\right|}
{\sqrt{\left\langle |f_1(k,\nu)|^2\right\rangle
\left\langle |f_2(k,\nu)|^2\right\rangle}}.
\end{equation}

\noindent A coherence value of 1 indicates a perfect linear relationship between the signals, whereas a value of 0 indicates no coherent relationship. The analysis focuses on phase shifts, $\delta\phi(k_h,\nu)$, obtained from  pairs of observables sampled at two atmospheric heights covering the photosphere. Throughout the analysis, the three-dimensional spectra are azimuthally averaged in the $(k_x,k_y)$ plane to obtain $(k_h,\nu)$ representations of both phase and coherence, where

\begin{equation}
k_h=\sqrt{k_x^2+k_y^2}.
\end{equation}

\noindent To aid the interpretation of the observed spectral signatures, we employ the standard dispersion relation for adiabatic acoustic--gravity waves under the Cowling approximation, which neglects perturbations of the gravitational potential \citep[][]{1981NASSP.450..263L}:

\begin{equation}
k_z^2=
\frac{(\omega^2-\omega_{ac}^2)(\omega^2-N^2)}
{c_s^2\omega^2}
-k_h^2,
\end{equation}

\noindent where $\omega=2\pi\nu$ is the angular frequency, $c_s$ is the sound speed, $\omega_{ac}$ is the acoustic cut-off frequency, and $N$ is the Brunt--Väisälä frequency. Regions with $k_z^2>0$ correspond to vertically propagating waves, whereas $k_z^2<0$ indicates evanescent behaviour.

\noindent Propagation boundaries in the $k_h-\nu$ space are determined from the condition $k_z^2=0$. These boundaries are calculated using atmospheric parameters from the MANCHA simulation evaluated at the upper formation height of each observable pair. The acoustic cut-off and Brunt--Väisälä frequencies are given by

\begin{equation}
\omega_{ac}^{2} = \frac{c_s^2}{4H_\rho^2}\left(1-2\frac{dH_\rho}{dz}
\right),
\end{equation}

\noindent and

\begin{equation}
N^2=\frac{g}{H_\rho}-\frac{g^2}{c_s^2},
\end{equation}

\noindent where $H_\rho$ is the density scale height and $g$ is the gravitational acceleration.\\

\noindent The condition $k_z^2=0$ yields two distinct propagation limits in the $k_h-\nu$ plane. The upper boundary is associated with the acoustic cut-off, while the lower boundary defines the propagation regime of atmospheric gravity waves and generally lies below the surface-gravity ($f$)-mode ridge. Since these limits vary with atmospheric height, the corresponding curves are displayed for the maximum upper height, i.e. $z=352$ km, used in our analysis. For reference, the dispersion relation of the surface-gravity wave is also shown in all phase and coherence diagrams.

\begin{figure*}
		\centering
	\includegraphics[width=0.97\linewidth]{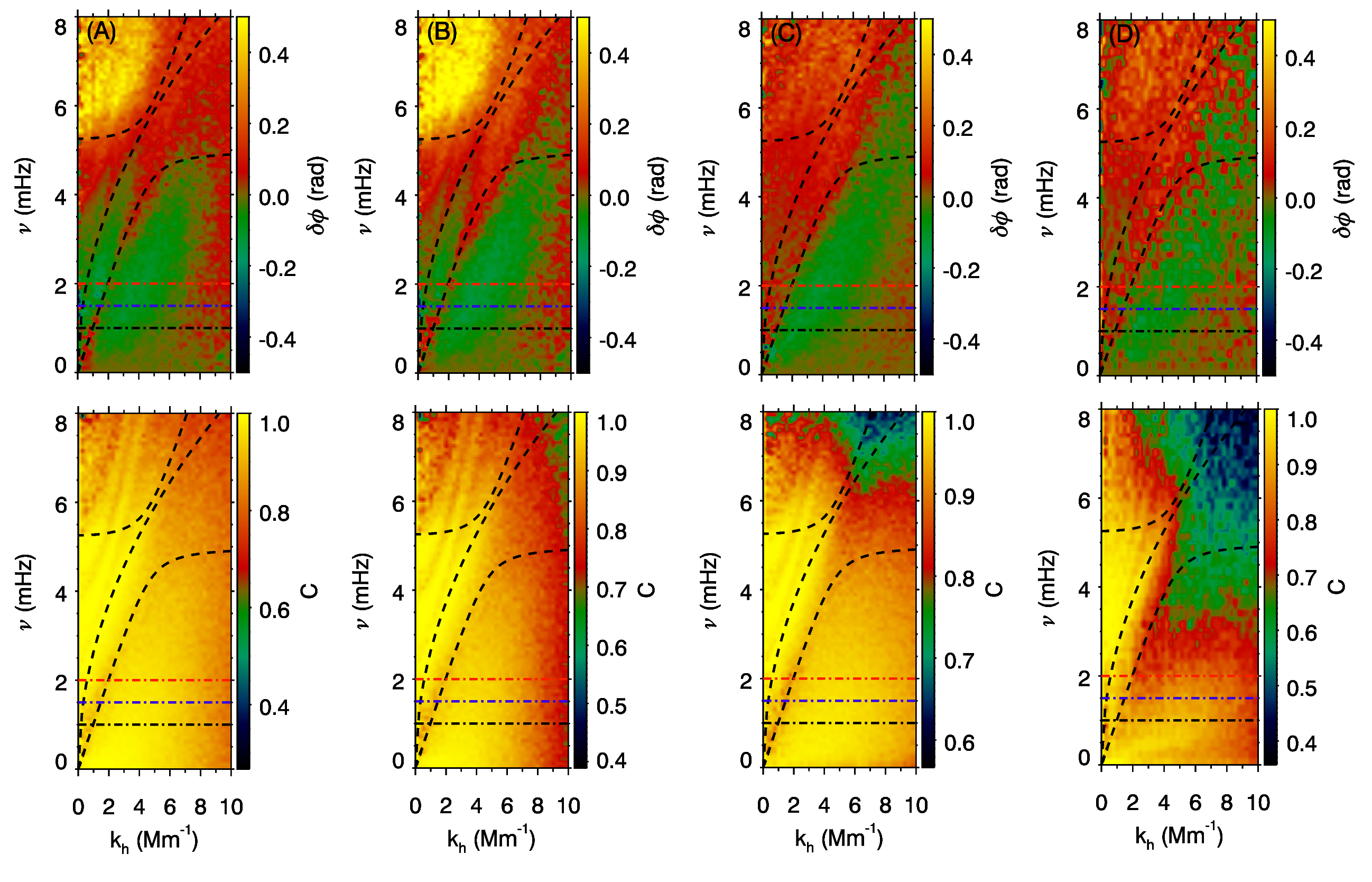}
\caption{Phase difference (top panels) and coherence (bottom panels) spectra between the LOS velocities at two different atmospheric layers. Each column corresponds to a different approach to determining the velocity. Column A: velocities at constant geometrical heights $z=32$ and $128$ km. Column B: velocities at constant optical depths $\log\tau=-0.9$ (mean height of approximately $39$ km) and $\log\tau=-1.6$ (mean height of approximately $144$ km). Column C: bisector-derived velocities from the synthetic Fe I 6173 {\AA} profile at the PV7 and PV0 levels. Column D: bisector-derived velocities from the SST observations of the same Fe I 6173 {\AA} line using the PV7--PV0 velocity pair. The black dashed curve separates the gravity-wave and acoustic-wave regimes. The black, blue, and red dash-dotted horizontal lines mark frequencies of 1.0, 1.5, and 2.0 mHz, respectively, selected for comparison of the phase differences and coherence across the different velocity diagnostics.}
	\label{K_omega_phase_diff}
\end{figure*}

\begin{figure*}
	\centering
	\includegraphics[width=0.97\linewidth]{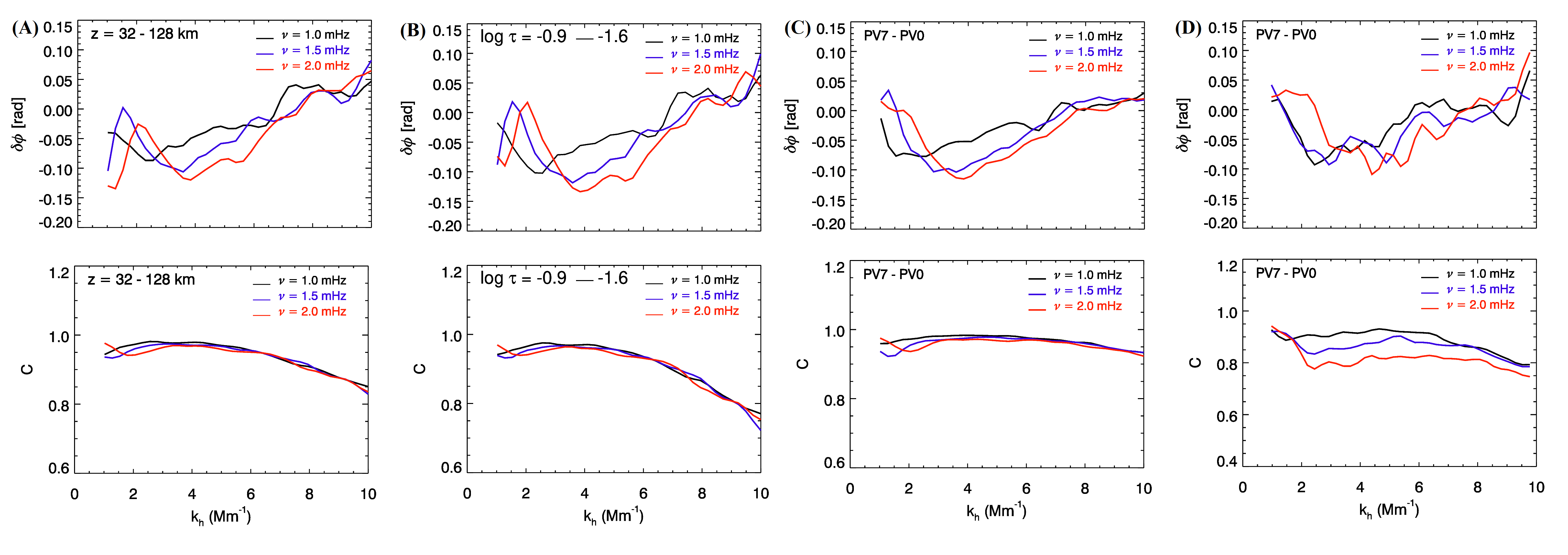}
\caption{The top panels show the phase difference ($\delta\phi(k_h)$) as a function of horizontal wavenumber ($k_h$) at the selected frequencies of 1.0, 1.5, and 2.0 mHz. Phase differences are shown for the four cases considered in Figure \ref{K_omega_phase_diff}: panel (A) between the LOS velocities at geometrical heights $z=32$ and $128$ km above $z=0$; panel (B) between the velocities at optical-depth levels corresponding to $\log\tau=-0.9$ and $-1.6$; panel (C) between the bisector-derived velocities PV7--PV0, corresponding to $\log\tau=-0.9$ and $-1.6$, respectively, from the synthetic Fe I 6173 {\AA} profile, providing a more direct comparison with the simulated velocities; and panel (D) between the corresponding bisector-derived velocities from the SST observations. The bottom panels show the corresponding coherence ($C(k_h)$) as a function of horizontal wavenumber for the same frequency cuts and pairs of heights or observables.}
	\label{Phase_coh_vlos_variation}
\end{figure*}

\subsection{\textbf{Velocity--velocity diagnostics}}

Figure \ref{K_omega_phase_diff} presents the $k_h-\nu$ phase difference and coherence diagnostic diagrams obtained from velocity--velocity pairs using four different velocity measurements. The first measurement consists of the true LOS velocities from the MANCHA simulation sampled at fixed geometrical heights $z=32$ and $128$ km. The second uses the corresponding velocity diagnostics sampled at fixed optical depths $\log\tau=-0.9$ and $-1.6$, which correspond to mean geometrical heights of approximately $39$ and $144$ km, respectively. These two measurements are obtained by sampling the same atmospheric model at constant geometrical heights and constant optical depths, respectively. The third measurement consists of multi-height LOS velocities derived using the bisector method from the synthetic Fe I 6173 {\AA} line profile, using the PV7 and PV0 levels. These bisector levels sample approximately the same range of optical depths as the fixed optical-depth velocity pairs (Figure \ref{Bis_optical_depth_cc}). The fourth measurement consists of bisector-derived multi-height LOS velocities obtained from the SST observations of the Fe I 6173 {\AA} line. We assume that the PV7--PV0 bisector levels in the SST observations sample approximately the same atmospheric layers as those in the synthetic profile. Consequently, the four diagnostics provide independent estimates of the velocity--velocity phase difference and coherence over approximately the same atmospheric height range. The sign convention is such that a positive phase difference indicates that the fluctuations in the upper layer lag behind those in the lower layer. Thus, negative phase differences in the gravity-wave regime provide the characteristic signature of upwardly propagating gravity waves, while positive phase differences in the acoustic regime are consistent with upwardly propagating acoustic waves.

To compare the wave signatures obtained from these four velocity measurements, we select three frequencies, $\nu=1.0$, $1.5$, and $2.0$ mHz, indicated by the black, blue, and red dash-dotted horizontal lines, respectively, in Figure \ref{K_omega_phase_diff}. Figure \ref{Phase_coh_vlos_variation} shows the corresponding variation of phase difference and coherence as a function of horizontal wavenumber $k_h$ at these three frequencies. For the comparison, we focus on the wavenumber interval $k_h = 2$--$4$ $\mathrm{Mm}^{-1}$. The lower limit is chosen such that the gravity-wave propagation boundary defined by the Brunt--Väisälä frequency reaches approximately $k_h = 2$ $\mathrm{Mm}^{-1}$ at $\nu = 2$ mHz, whereas the upper limit of $k_h = 4$ $\mathrm{Mm}^{-1}$ is selected because the coherence decreases significantly beyond this value. The cuts are shown for all four velocity--velocity diagnostics, allowing a direct comparison between the geometrical-height velocities, optical-depth velocities, synthetic bisector-derived velocities, and bisector-derived velocities from the SST observations.

The phase-difference spectra show negative phase shifts in the gravity-wave regime for all four velocity measurements, accompanied by relatively high coherence. This behaviour is consistently observed at the selected frequencies of 1.0, 1.5, and 2.0 mHz. Importantly, all four velocity measurements exhibit remarkably similar phase-difference and coherence patterns across the examined horizontal-wavenumber range. The velocity pairs based on fixed geometrical heights and optical depths show closely matching behaviour, while the bisector-derived velocities from both the synthetic Fe I 6173 {\AA} profile and the SST observations reproduce similar characteristic phase shifts and coherence patterns. Thus, despite the different methods used to obtain the four velocity measurements, they yield consistent wave-propagation signatures.

The agreement among the four diagnostics is particularly evident in the gravity-wave regime, where negative phase differences are accompanied by relatively high coherence. This consistency indicates that the characteristic gravity-wave signature is preserved when moving from the simulation velocities sampled at fixed geometrical heights to those sampled at constant optical depths and, further, to bisector-derived velocities from synthetic and observed Fe I 6173 {\AA} profiles.

\subsection{\textbf{Temperature--temperature and intensity--intensity diagnostics}}

\begin{figure}
	\centering
	\includegraphics[width=0.99\linewidth]{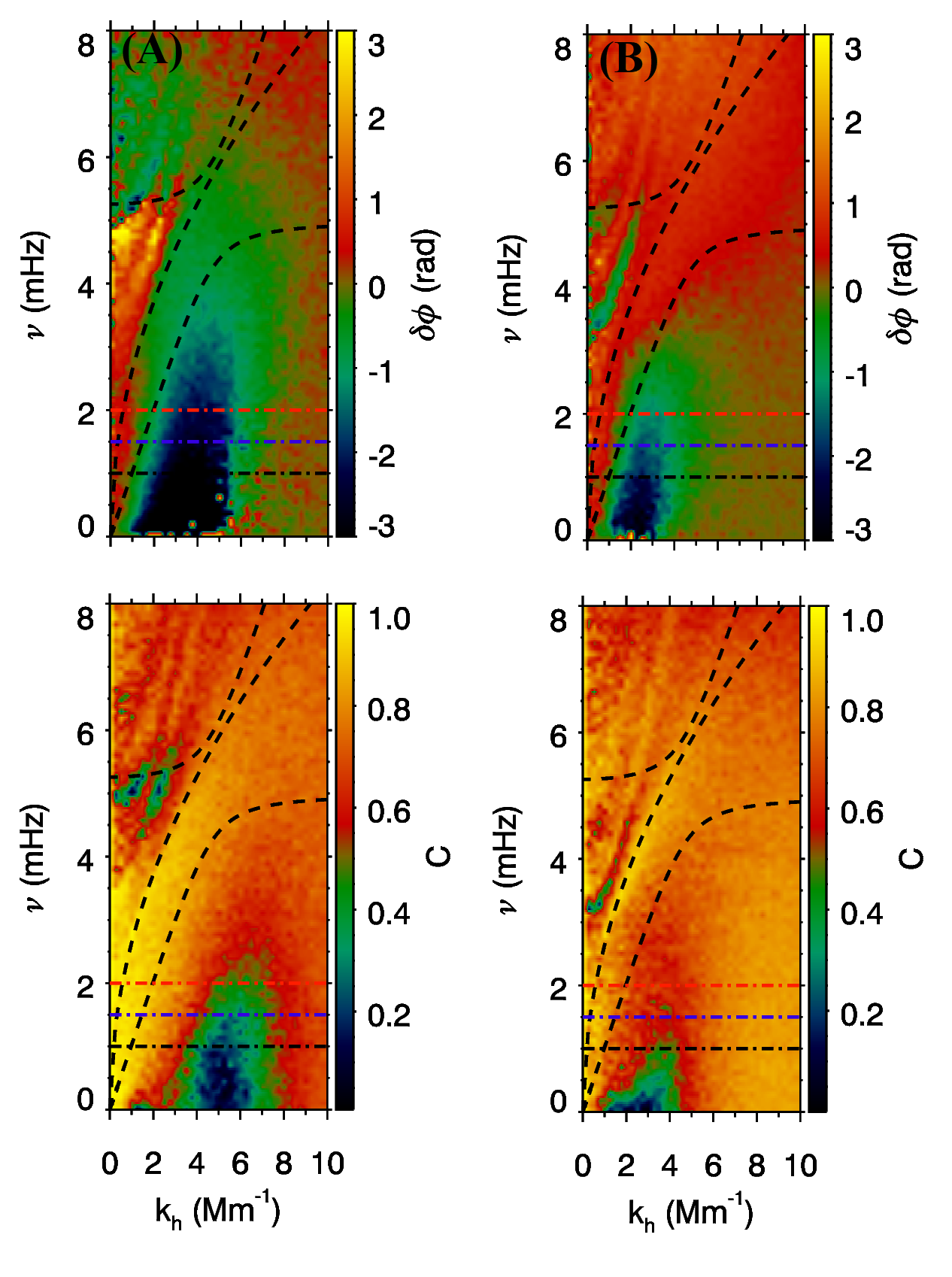}
    	\caption{The top panels show the phase difference diagrams, derived from temperature--temperature measurements between the geometrical heights $z = 32$ and $128$ km above $z = 0$ (panel A), and between constant optical-depth surfaces corresponding to $\log \tau = -0.9$ and $-1.6$ (panel B). The bottom panels show the corresponding coherence ($C$) spectra for the same pairs of sampling layers used in the phase-difference analysis.}
	\label{k-omega-phase-coh-Temp-14-0916}
\end{figure}

Figure \ref{k-omega-phase-coh-Temp-14-0916} shows the phase difference and coherence diagnostic diagrams derived from two-height temperature--temperature measurements, with the same MANCHA atmosphere sampled at constant geometrical heights ($z = 32$--$128$ $\mathrm{km}$) and constant optical depths ($\log \tau = -0.9$ to $-1.6$). In contrast to the velocity--velocity diagnostics discussed above (panels (A) and (B) of Figure \ref{K_omega_phase_diff}), the temperature-based measurements reveal pronounced differences between the two sampling approaches. These differences are evident not only in the magnitude of the phase shifts but also in the spatial extent of the gravity-wave signatures across the $k$--$\omega$ space within the gravity-wave regime boundary. The phase-difference diagrams (top panels of Figure \ref{k-omega-phase-coh-Temp-14-0916}) show that the negative phase shifts associated with upwardly propagating gravity waves occupy a substantially larger region of the $k$--$\omega$ space when the measurements are obtained at constant geometrical heights (panel A). In contrast, when the same atmosphere is sampled at constant optical depths, the gravity-wave signature is confined to a smaller region and exhibits systematically smaller phase-shift amplitudes (panel B).

This behavior indicates that temperature fluctuations are strongly influenced by opacity effects, which modify the apparent wave properties when sampled within constant optical depth. As shown by the temperature variation in Figure \ref{mancha_temp_variation}, the atmosphere has a strong temperature gradient over the height range considered here. At fixed optical depth, opacity variations can shift the geometrical height of the sampled layers, causing the measured temperature signal to be affected by the background temperature gradient. Consequently, the temperature fluctuations sampled at constant optical depth may contain contributions from both the intrinsic thermal perturbations and the displacement of the optical-depth surfaces through the background temperature stratification. This effect can therefore modify the measured phase and amplitude of the temperature fluctuations, resulting in a weaker and more spatially confined over $k-\omega$ space, gravity-wave signature compared with that obtained at fixed geometrical heights.


\begin{figure}
	\centering
	\includegraphics[width=0.99\linewidth]{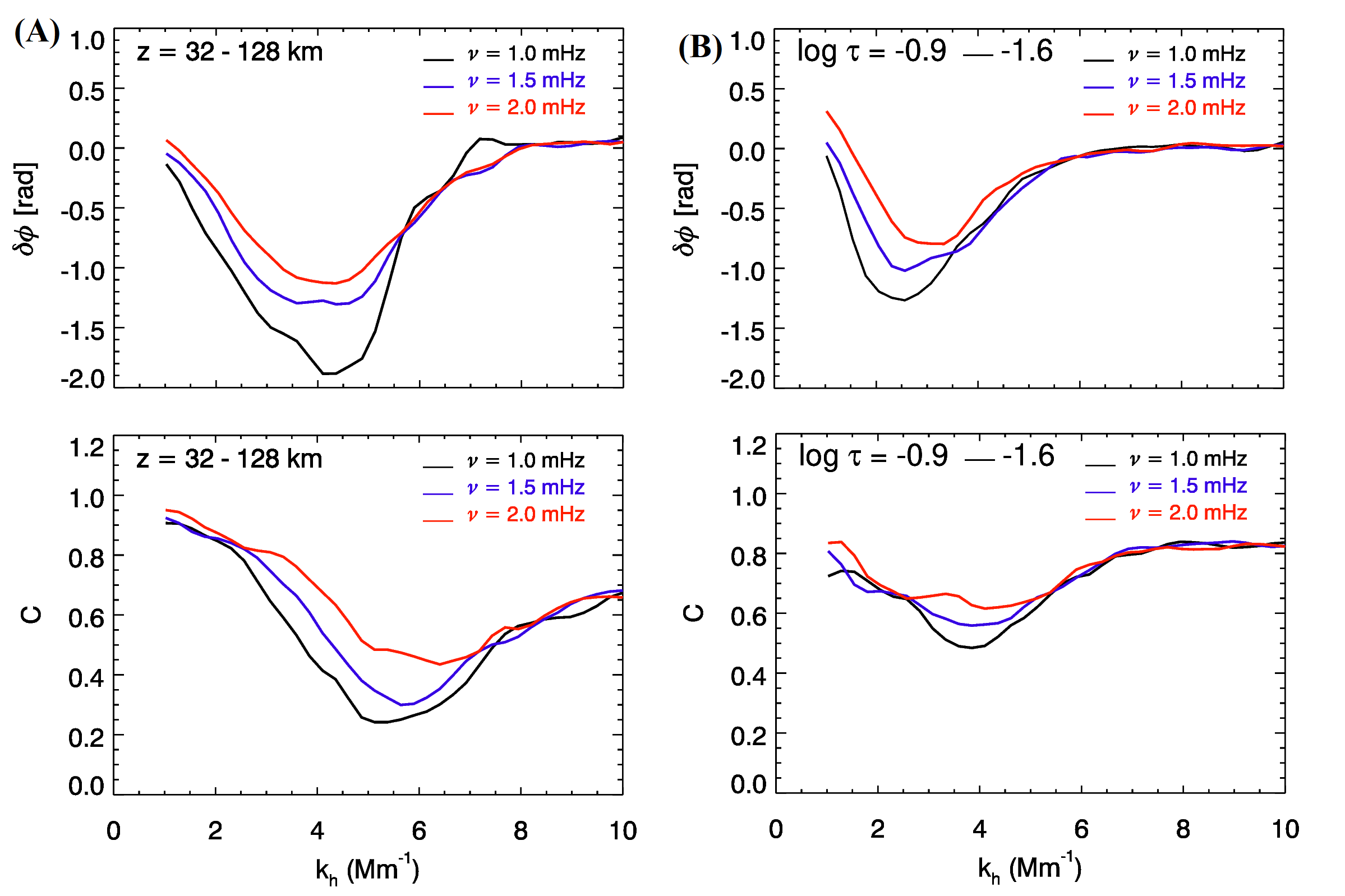}
    	\caption{The top panels show the phase difference ($\delta \phi (\nu)$) as a function of horizontal wavenumber ($k_{h}$) at frequencies of 1.0, 1.5, and 2.0 mHz, derived from temperature--temperature measurements between geometrical heights $z = 32$ and $128$ km above $z = 0$ (panel A), and between constant optical-depth surfaces corresponding to $\log \tau = -0.9$ and $-1.6$ (panel B), as shown in Figure \ref{k-omega-phase-coh-Temp-14-0916}. The bottom panels show the corresponding coherence ($C$) spectra for the same sampling-layer pairs used in the phase-difference analysis.}
	\label{Phase_coh_temp_variation_14_0916}
\end{figure}


A more quantitative comparison is provided in Figure \ref{Phase_coh_temp_variation_14_0916}, which shows the phase differences and coherence as functions of horizontal wavenumber for the selected frequencies $\nu = 1.0$, $1.5$, and $2.0$ mHz. Within the wavenumber range $k_{h} = 2$--$4$ $\mathrm{Mm}^{-1}$, where the coherence remains relatively high ($C \gtrsim 0.5$), significant differences are found between the phase shifts inferred from the two sampling approaches. For example, at $k_{h} \approx 3$ $\mathrm{Mm}^{-1}$, the phase difference measured between the geometrical heights $z = 32$ and $128$ km varies from approximately $-1.0$ to $-2.0$ rad, depending on frequency (top-left panel (A) of Figure \ref{Phase_coh_temp_variation_14_0916}). In contrast, the corresponding phase differences derived from the same atmosphere sampled at constant optical depths are considerably smaller, typically ranging between $-0.6$ and $-1.2$ rad (top-right panel (B) of Figure \ref{Phase_coh_temp_variation_14_0916}). Thus, the temperature diagnostics sampled at constant geometrical heights yield phase shifts that are systematically larger by several tenths of a radian throughout the gravity-wave regime. The coherence values obtained using the two sampling approaches are broadly comparable, particularly within the wavenumber range $k_{h} = 2$--$4$ $\mathrm{Mm}^{-1}$ used for the phase-shift comparison. This similarity in coherence suggests that the differences in phase shifts are not simply associated with a loss of coherence between the signals. Rather, they indicate that the inferred temperature phase differences depend on the choice of sampling surface. The differences are consistent with temperature-based diagnostics being more sensitive to opacity-related effects than the velocity-based diagnostics discussed above. Consequently, the inferred gravity-wave properties can depend systematically on whether the temperature signals are sampled at constant geometrical heights or optical depths.

The same trend is observed when the height separation between the two temperature measurements is increased. Figures \ref{k-omega-phase-coh-Temp-111-0931} and \ref{Phase_coh_temp_variation_111_0931} show the corresponding phase-difference and coherence diagnostics for a larger separation between the sampling layers. As in the case with the smaller separation, the diagnostics sampled at constant geometrical heights exhibit stronger and more extended gravity-wave phase shifts than those sampled at constant optical depths. Although the absolute phase shifts increase with increasing  separation between the sampling layers, the systematic discrepancy between the two sampling approaches persists. This further supports the conclusion that the inferred temperature-based phase differences are sensitive to the choice of sampling surface and may be affected by opacity variations.

\begin{figure}
		\centering
	\includegraphics[width=0.99\linewidth]{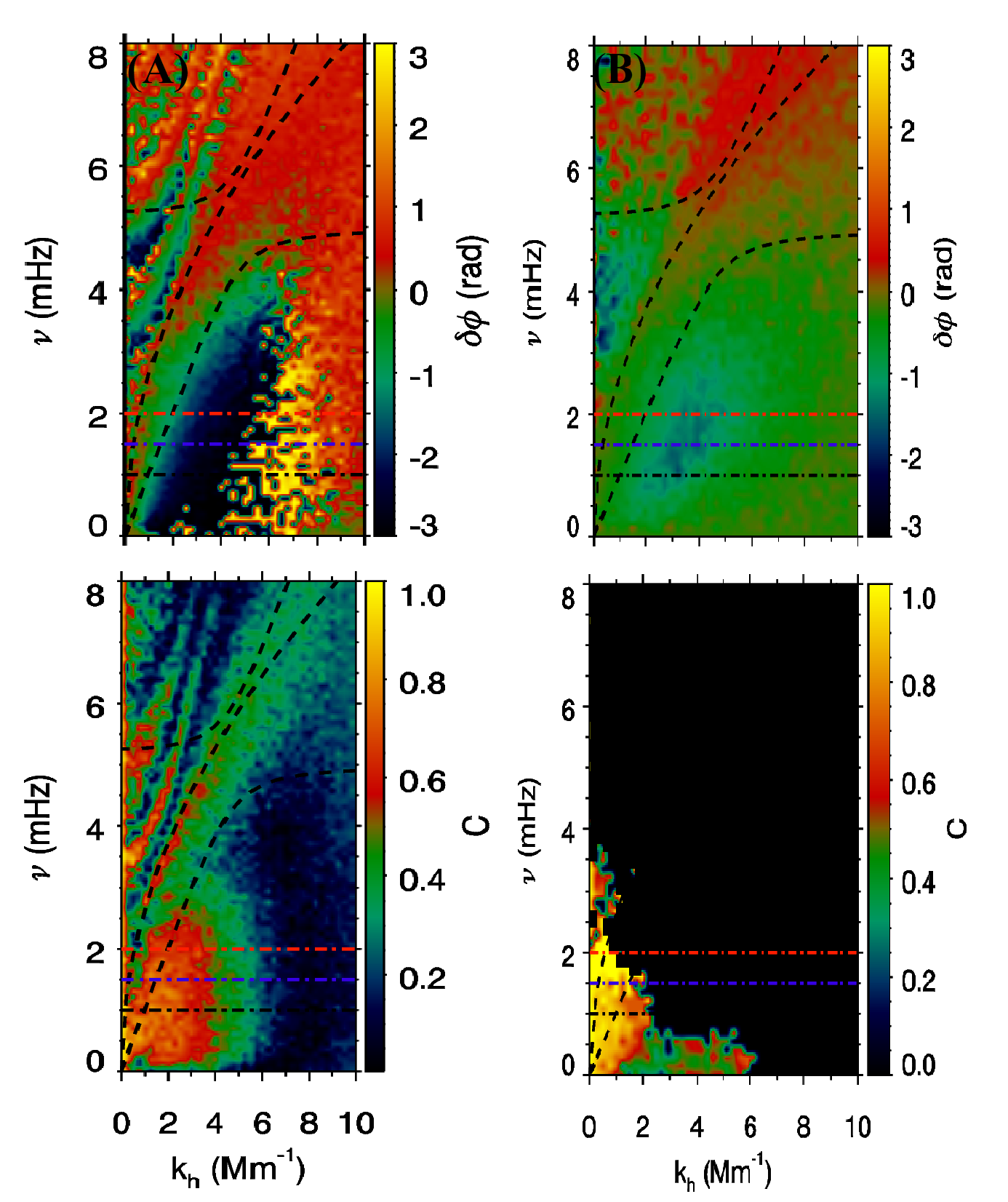}
	\caption{The top panels show the phase-difference diagrams derived from the continuum intensity--line-core intensity ($I_{\rm c}$--$I_{\rm core}$) pair obtained from the synthetic Fe I 6173 {\AA} line profiles generated using the MANCHA simulation (panel A) and from the SST observations (panel B). The bottom panels show the corresponding coherence diagrams for the same observables used to estimate the phase differences.}
	\label{K_omega_phase_diff_Ic_Icore}
\end{figure}

\begin{figure}
	\centering
	\includegraphics[width=0.99\linewidth]{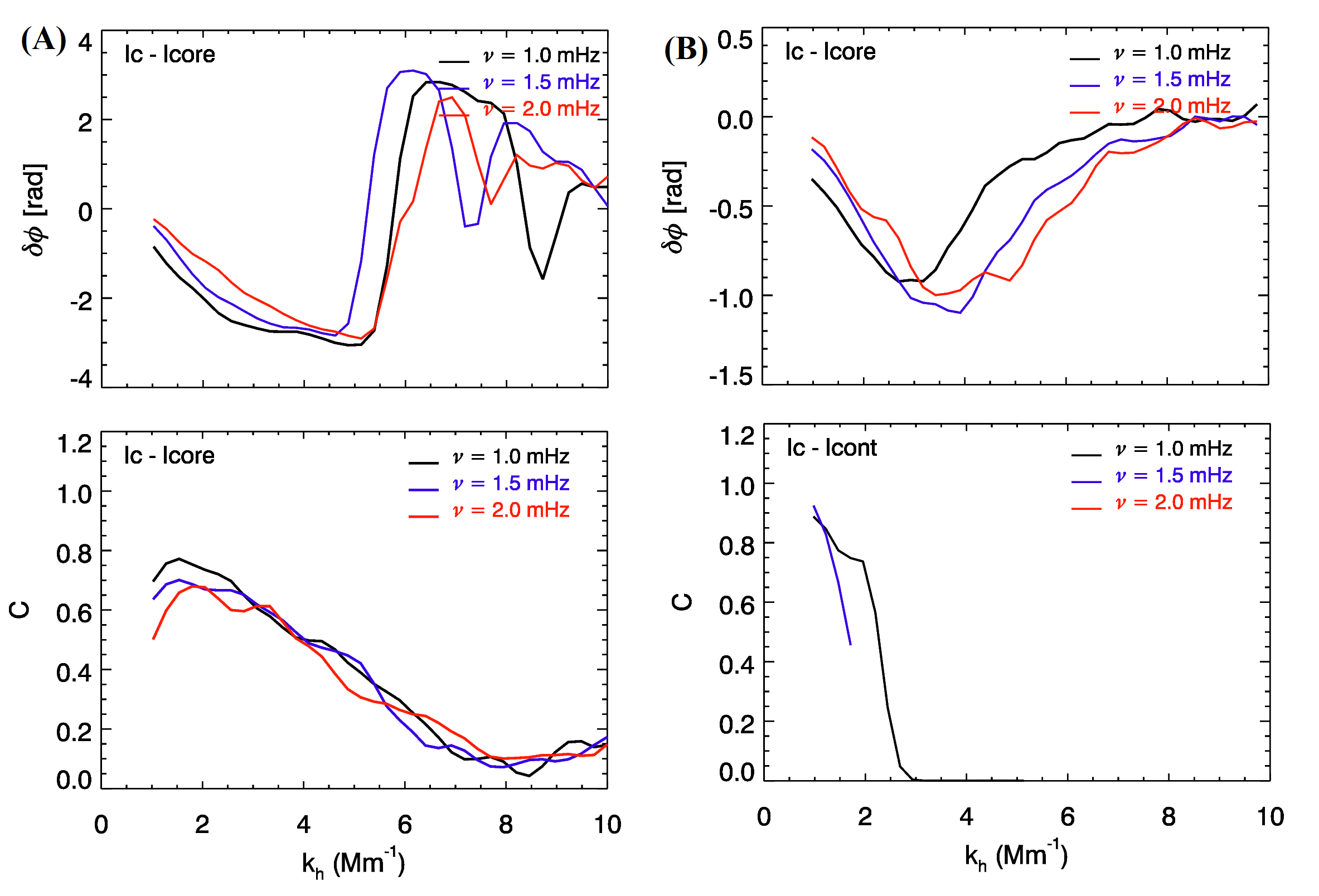}
    	\caption{The top panels show the phase difference ($\delta \phi (\nu)$) as a function of horizontal wavenumber ($k_{h}$) at frequencies of 1.0, 1.5, and 2.0 mHz, derived from the continuum intensity and line-core intensity ($I_{\rm c}$--$I_{\rm core}$) of the synthetic Fe I 6173 {\AA} line profiles generated using the MANCHA simulation (panel A) and from the SST observations (panel B), respectively, as shown in Figure \ref{K_omega_phase_diff_Ic_Icore}. The bottom panels show the corresponding coherence ($C$) spectra for the same observables used in the phase-difference analysis.}
	\label{Phase_coh_Ic_Icore_variation}
\end{figure}

To further examine whether the gravity-wave signatures identified in the thermodynamic variables are also detectable in spectroscopic observables, we first performed the phase-difference and coherence analysis using the synthetic Fe I 6173 {\AA} intensity profiles. We considered the continuum intensity ($I_{\rm c}$), which predominantly samples the lower photosphere, and the line-core intensity ($I_{\rm core}$), which samples the middle photospheric layers. The resulting intensity--intensity ($I_{\rm c}$--$I_{\rm core}$) phase-difference and coherence diagrams are presented in column (A) of Figure \ref{K_omega_phase_diff_Ic_Icore}. The diagnostics show negative phase differences in the gravity-wave regime, indicating the presence of the characteristic gravity-wave signature in the synthetic spectroscopic observables. We then applied the same $I_{\rm c}$--$I_{\rm core}$ analysis to the SST observations of the Fe I 6173 {\AA} line. The resulting $k$--$\omega$ phase-difference and coherence diagrams are shown in column (B) of Figure \ref{K_omega_phase_diff_Ic_Icore}. The SST observations also show negative phase differences in the gravity-wave regime, although the phase shifts are generally smaller than those obtained from the synthetic spectra, and the region of high coherence is restricted to a smaller part of the $k$--$\omega$ diagram. The corresponding phase-difference and coherence variations at $\nu = 1.0$, $1.5$, and $2.0$ mHz are shown in columns (A) and (B) of Figure \ref{Phase_coh_Ic_Icore_variation}. Overall, the synthetic and observed $I_{\rm c}$--$I_{\rm core}$ diagnostics exhibit the same qualitative gravity-wave signature, with some quantitative differences in the phase shifts and coherence. This indicates that intensity-based diagnostics can retain detectable gravity-wave signatures, although their measured phase differences and coherence may differ quantitatively between synthetic and observed data. Moreover, because the continuum and line-core intensities are influenced by radiative transfer and opacity, their phase differences should not necessarily be interpreted as the phase shifts between waves sampled at two fixed geometrical heights.

\section{Discussions and conclusions}
\label{Sect4}

In this work, we investigated how the choice of observable and atmospheric sampling affects the detection and interpretation of solar atmospheric gravity waves. We combined realistic radiative-hydrodynamic simulations from the MANCHA code with synthetic Fe I 6173 {\AA} spectra and high-resolution SST observations. In particular, we examined whether the phase propagation signatures inferred from Doppler velocity and intensity diagnostics are preserved when the atmospheric signals are sampled at constant geometrical height, constant optical depth, or through spectroscopic observables. 

\cite{2021RSPTA.37900170F} performed a comparative investigation of gravity waves using the MANCHA, CO5BOLD, Bifrost, and MURaM simulations and found broadly consistent gravity-wave signatures, with differences related to the individual simulation setups and physical assumptions. Our velocity--velocity analysis further supports the robustness of velocity as a diagnostic of gravity-wave propagation. The phase-difference and coherence spectra obtained from velocities sampled at fixed geometrical heights and fixed optical depths show very similar behaviour over the horizontal-wavenumber range considered here. This similarity is also recovered when the velocity is inferred from the synthetic Fe I 6173 {\AA} line using the bisector method and is qualitatively consistent with the bisector-derived velocities from the SST observations. Thus, despite the different sampling procedures, the characteristic phase relationship associated with gravity-wave propagation is preserved in the velocity diagnostics. This suggests that, over the atmospheric height range examined here, the velocity-based phase diagnostics are relatively insensitive to whether the velocity is sampled on a geometrical-height or optical-depth scale. These results also indicate that spectroscopic Doppler-velocity measurements can capture the characteristic signatures of gravity waves. 

The temperature diagnostics show a different behaviour. The phase-difference spectra obtained from fixed geometrical heights exhibit a more extended gravity-wave signature and larger phase shifts than those obtained from fixed optical depths. Here, a more extended signature refers to the larger region of the $k_h$--$\nu$ diagram over which the negative phase differences associated with upwardly propagating gravity waves are present. Conversely, a more confined signature refers to a smaller region of this diagram in which these phase differences are detected. This difference is related to the sensitivity of optical depth to the local thermodynamic conditions. Variations in opacity can shift constant optical-depth surfaces in geometrical height, and these shifts become important in the presence of a vertical temperature gradient. As a result, temperature fluctuations sampled at constant optical depth can contain contributions from both the intrinsic temperature perturbation and the displacement of the sampling layer. The phase difference obtained from such measurements therefore does not represent simply the phase difference between temperature perturbations at two fixed geometrical heights. This provides a possible explanation for the smaller phase-shift amplitudes and the more spatially confined gravity-wave signatures found in the temperature diagnostics sampled at constant optical depth.

The synthetic Fe I 6173 {\AA} intensity diagnostics provide an important link between the simulation variables and spectroscopic observables. The $I_{\rm c}$--$I_{\rm core}$ phase-difference spectra retain the characteristic gravity-wave signature, but their phase shifts are smaller than those obtained from the temperature diagnostics at fixed geometrical heights. This correspondence is consistent with the fact that spectroscopic intensities are formed according to the radiative-transfer conditions of the atmosphere rather than at fixed geometrical heights. Therefore, the $I_{\rm c}$--$I_{\rm core}$ phase difference should not be interpreted as a direct measurement of the phase difference between temperature perturbations at two fixed heights. Instead, it reflects the combined effects of atmospheric fluctuations, opacity, radiative transfer, and the thermodynamically dependent formation of the continuum and line-core signals. Thus, although intensity diagnostics can reveal the presence of gravity-wave signatures, their phase differences do not necessarily provide a direct measure of the phase propagation between two fixed geometrical heights.

The SST observations show the same qualitative gravity-wave signature as the synthetic spectroscopic diagnostics, although quantitative differences are present in both phase difference and coherence. In particular, the observed $I_{\rm c}$--$I_{\rm core}$ phase shifts are generally smaller than those obtained from the fixed-geometrical-height temperature diagnostics, and the synthetic intensity diagnostics are broadly consistent with the optical-depth temperature behaviour. This agreement provides observational support for the interpretation that the spectroscopic intensity diagnostics are sensitive to the optical-depth structure of the atmosphere. At the same time, the differences between the synthetic and observed diagnostics indicate that additional effects associated with the observed atmospheric structure, radiative transfer, instrumental response, and effective formation heights can influence the measured phase relationships. The regions of very low coherence found in parts of the SST $I_{\rm c}$--$I_{\rm core}$ spectra further indicate that the continuum and line-core intensity fluctuations are not uniformly correlated over the entire $k_h$--$\nu$ domain.

An additional consideration is the possible influence of magnetic fields on the observed diagnostics. The MANCHA simulations used in this study are field-free, whereas the SST observations correspond to relatively quiet-Sun regions dominated by the magnetic network. Although these regions do not contain the strong magnetic structures, such as plages, pores, or sunspots, discussed in the context of the magnetic influence on gravity waves, magnetic effects cannot be completely excluded. This is particularly relevant for the Fe I 6173 {\AA} line, which is magnetically sensitive and has an effective Landé factor of $g_{\rm eff}=2.5$. Consequently, magnetic fields may contribute to the quantitative differences between the synthetic and observed continuum-to-line-core intensity diagnostics. Atmospheric seeing and associated observational effects may also contribute to these differences. We note that the bisector-derived velocities from the observed Fe I 6173 {\AA} line show only minor differences compared with the simulated velocities and bisector-derived velocities from synthetic line. This suggests that the combined effects of magnetic fields and other observational factors do not introduce substantial discrepancies in the velocity-based diagnostics considered here. However, this agreement does not rule out possible contributions to the intensity-based diagnostics.

All together, the results show a clear distinction between velocity and temperature/intensity diagnostics. The velocity--velocity diagnostics give consistent phase-difference and coherence signatures across geometrical-height, optical-depth, synthetic bisector, and observed bisector velocities. In contrast, the temperature diagnostics are more sensitive to the choice of atmospheric coordinate, while the spectroscopic intensity diagnostics show behaviour more closely related to the optical-depth temperature diagnostics. This distinction is important when interpreting observational phase differences, since an observed phase shift between two intensity signals cannot be directly equated with the phase shift between two fixed geometrical layers.

Our results are broadly consistent with previous observational studies that have identified gravity-wave signatures using ultraviolet and continuum intensity measurements \citep{2001A&A...379.1052K,2003A&A...407..735R,2023AdSpR..72.1898K}. While these studies demonstrate that intensity observations can be used to identify gravity-wave signatures, our results emphasize that the quantitative interpretation of their phase differences requires consideration of radiative-transfer effects and the effective formation heights of the observables. This is also relevant to estimates of gravity-wave energetics. Measurements of the vertical energy flux depend on the inferred wave energy density and group velocity \citep{2008ApJ...681L.125S}. The group velocity is determined from the wave dispersion relation, which involves the phase velocity. The phase velocity is commonly inferred from phase differences between signals measured at different atmospheric heights. Therefore, if phase differences derived from intensity or temperature diagnostics are interpreted as representing waves propagating between two fixed geometrical heights, the resulting phase velocity and group velocity may be overestimated. This can, in turn, lead to an overestimation of the inferred wave energy flux compared with estimates based on velocity--velocity diagnostics. Here, we found that temperature--temperature diagnostics at constant optical depth underestimate the phase differences, which would lead to an overestimation of the phase velocity. We emphasize that the present study does not quantitatively reassess previously reported energy-flux estimates, including those of \cite{2008ApJ...681L.125S}. Rather, our results identify a potential systematic source of overestimation when intensity- or temperature-based phase differences are used to infer wave-propagation properties and estimate energy fluxes. Future observational studies should therefore account for the distinction between diagnostics sampled at constant geometrical heights and optical depths, preferably by combining Doppler-velocity measurements with intensity diagnostics and forward modelling of the spectral profiles. 

In conclusion, our analysis shows that the qualitative detection of solar atmospheric gravity waves is relatively robust in velocity diagnostics but is more sensitive to atmospheric sampling in temperature and intensity diagnostics. Overall, the qualitative agreement between the numerical simulations, forward-synthesized spectra, and SST observations supports the potential of the MANCHA code to examine the connection between atmospheric gravity-wave properties and their spectroscopic signatures. The results also suggest that opacity, formation-height variations, and radiative-transfer effects should be considered when interpreting phase diagnostics derived from spectroscopic intensity measurements.

\begin{acknowledgements}

Financial support from grants PID2024-156538NB-I00, funded by MCIN/AEI/ 10.13039/501100011033 and by “ERDF A way of making Europe”, and from grant CNS2023-145233 funded by MICIU/AEI/10.13039/501100011033 and by “European Union NextGeneration EU/PRTR” is gratefully acknowledged. TF acknowledges grant RYC2020-030307-I funded by MCIN/AEI/ 10.13039/501100011033 and by “ESF Investing in your future”. This work is part of grant CEX2025-001609-S, awarded to the Instituto de Astrof\'isica de Canarias under the Severo Ochoa Centre of Excellence program and funded by MICIU/AEI/10.13039/501100011033. The Swedish 1-m Solar Telescope is operated on the island of La Palma by the Institute for Solar Physics of Stockholm University in the Spanish Observatorio del Roque de los Muchachos of the Instituto de Astrof\'isica de Canarias. The Swedish 1-m Solar Telescope, SST, is co-funded by the Swedish Research Council as a national research infrastructure (registration number 4.3-2021-00169). The authors thankfully acknowledge the technical expertise and assistance provided by the Spanish Supercomputing Network (Red Española de Supercomputación), as well as the computer resources used: the LaPalma Supercomputer, located at the Instituto de Astrofísica de Canarias.

\end{acknowledgements}

%
%
\bibliographystyle{aa}       
\bibliography{reference.bib}    

\clearpage
\appendix
\section{Additional Figures}
\label{app:phase}

\begin{figure}[h!]
	\centering
	\includegraphics[width=0.98\linewidth]{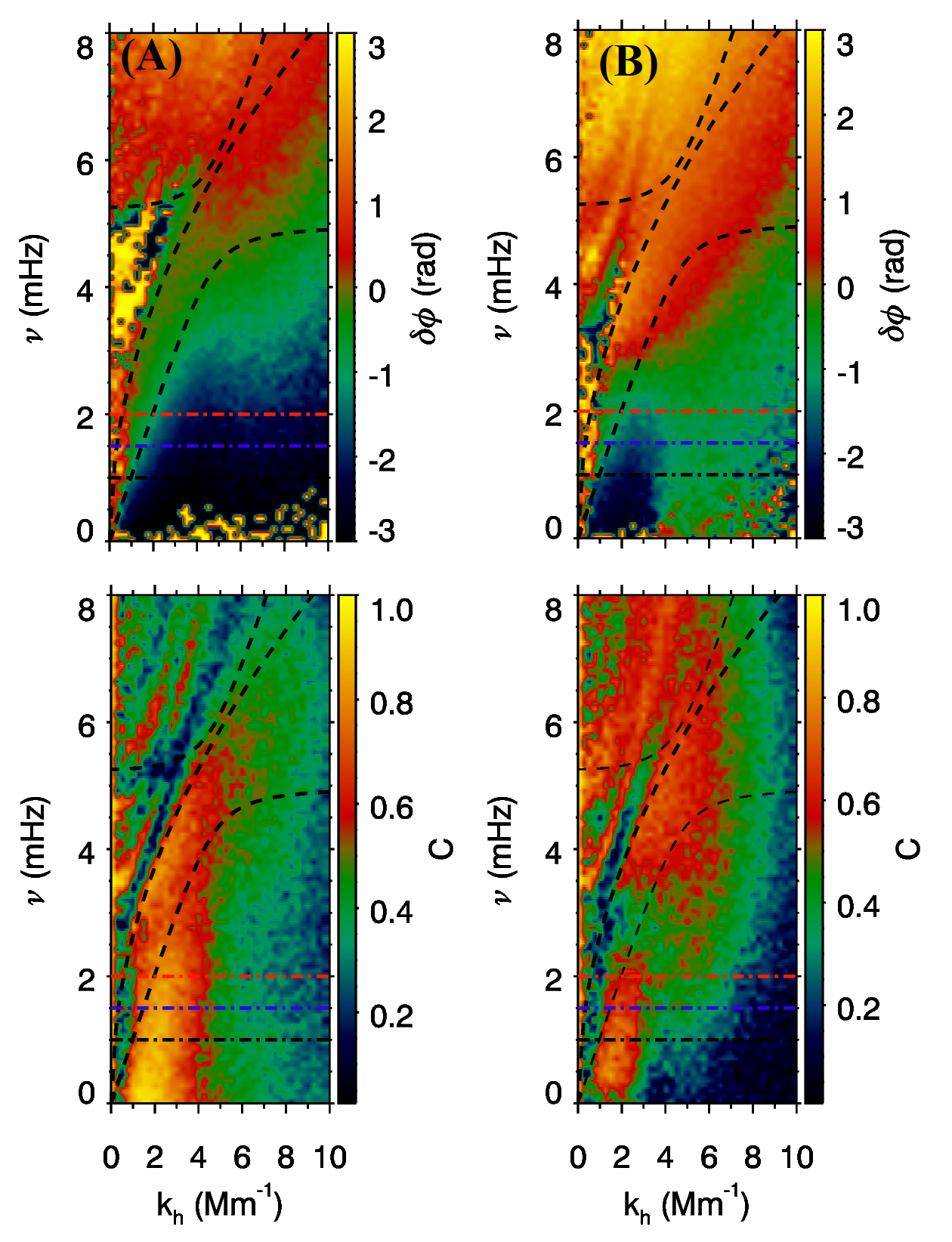}
  	\caption{The top panels show the phase difference diagrams, derived from the temperature--temperature between geometrical heights $z = 32$ and $352$ km above $z = 0$ (panel A), and between the optical-depth layers corresponding to $\log \tau = -0.9$ and $-3.2$ (panel B). The bottom panels show the corresponding coherence (C) spectra for the same height pairs used in the phase-difference analysis.}
	\label{k-omega-phase-coh-Temp-111-0931}
\end{figure}
\FloatBarrier

\begin{figure}[h!]
	\centering
	\includegraphics[width=0.98\linewidth]{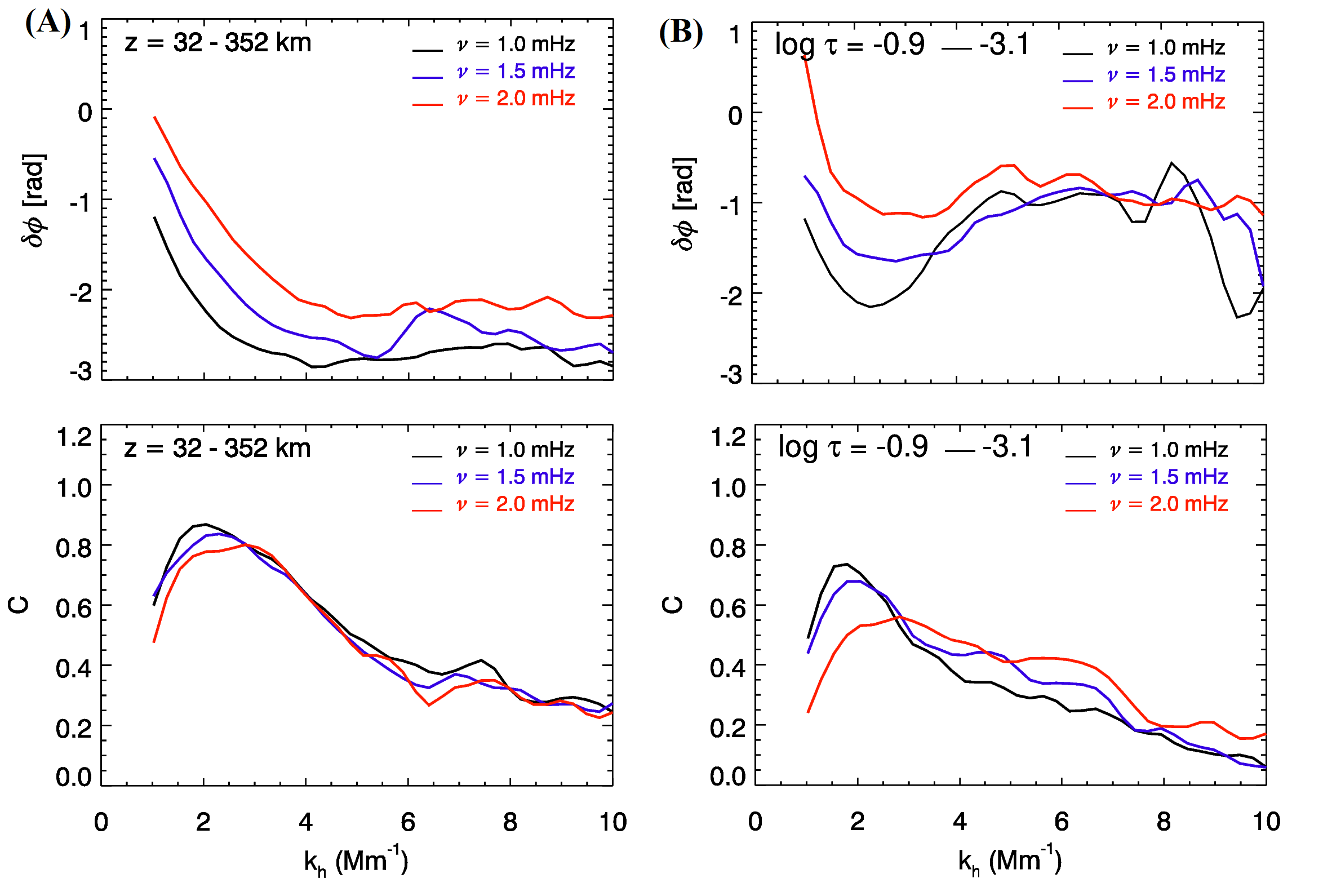}
	    	\caption{The top panels show the phase difference ($\delta \phi (\nu)$) as a function of wavenumber ($k_{h}$) at frequencies of 1.0, 1.5, and 2.0 mHz, derived from the temperature--temperature between geometrical heights $z = 32$ and $352$ km above $z = 0$ (panel A), and between the optical-depth layers corresponding to $\log \tau = -0.9$ and $-3.2$ (panel B), as shown in the Figure \ref{k-omega-phase-coh-Temp-111-0931}. The bottom panels show the corresponding coherence (C) spectra for the same height pairs used in the phase-difference analysis.}
	\label{Phase_coh_temp_variation_111_0931}
\end{figure}
\FloatBarrier

\begin{figure}[h!]
	\centering
	\includegraphics[width=0.99\linewidth]{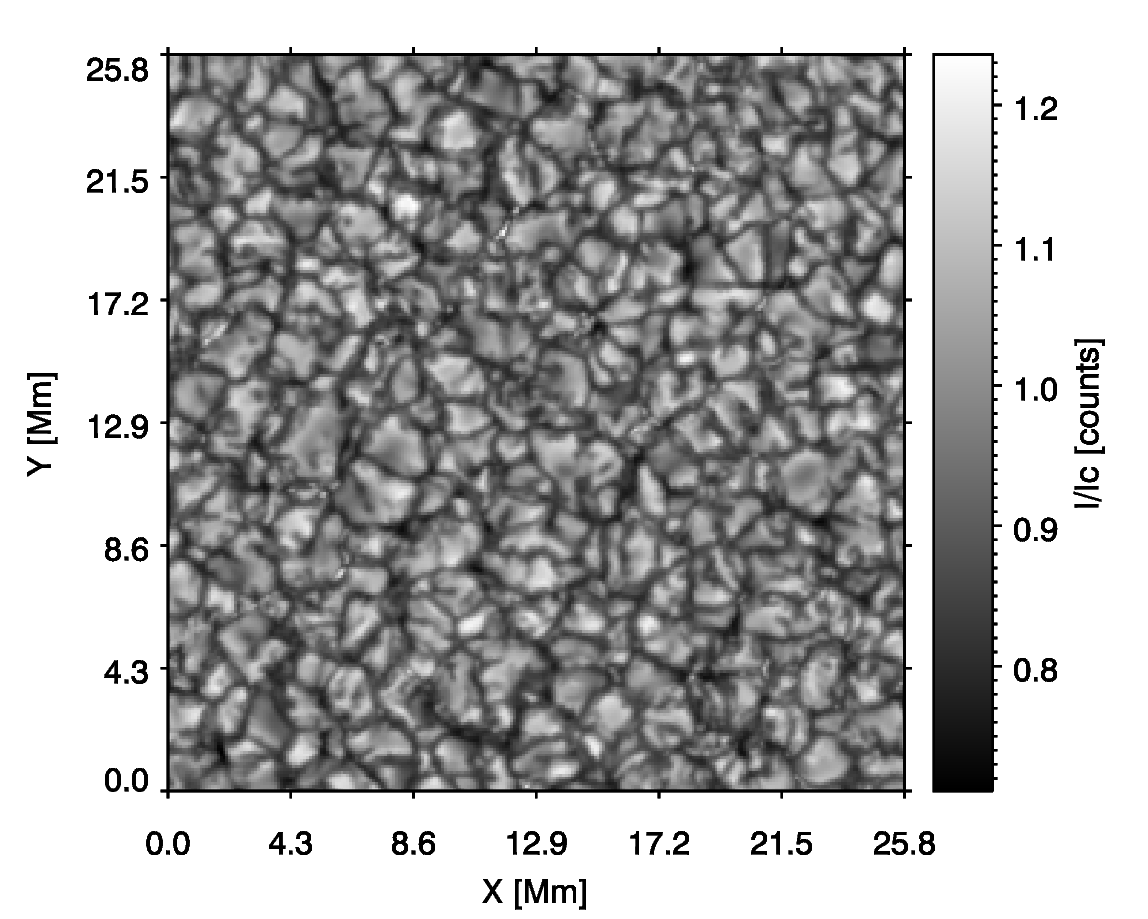}
	\caption{Sample photospheric continuum intensity map obtained from the SST observations.}
	\label{Sample_maps_Ic_SST_appendix}
\end{figure}
\FloatBarrier

\end{document}